\documentclass[a4paper,11pt,normalheadings,abstracton]{scrartcl}
\usepackage{amsmath}
\usepackage{amssymb}
\usepackage{amsfonts}
\usepackage{amscd}
\usepackage{bbm} 
\usepackage{fleqn} 
\usepackage[footnotesize]{caption}
\usepackage{graphicx} 
\usepackage{braket} 
\usepackage{mathrsfs}
\usepackage[center,footnotesize,hang]{subfigure}
\usepackage[bbgreekl]{MyBbol}
\usepackage{accents}
\usepackage{BetterCite}

\allowdisplaybreaks[1] 

\newcommand{\CenterEps}[2][1]{
\ensuremath{\vcenter{\hbox{\includegraphics[scale=#1]{#2.eps}}}}
}

\newcommand{\RaiseBrace}[1]{\raise3pt\hbox{$\displaystyle#1$}}

\def\D{\mathrm{d}}

\def\Tr{\text{Tr}}

\def\<{\left\langle}
\def\>{\right\rangle}

\newcommand{\SuperField}[1]{\bbsymbol{#1}}

\begin{document}


\begin{flushright}
{\small TUM-HEP-476/02}
\end{flushright}

\begin{center}
\sffamily\bfseries\Large
Renormalization Group Analysis of Neutrino 
Mass Parameters\footnote{%
Talk given at 
the 10th International Conference on 
Supersymmetry and Unification of 
Fundamental Interactions  (SUSY02), June 17 -- 23, DESY Hamburg.
Based on collaborations with
Manuel Drees, J\"orn Kersten, Manfred Lindner and Michael Ratz.
} \\
\end{center}

\begin{center}
{\large 
Stefan Antusch\footnote{{\em E-mail:} \texttt{santusch@ph.tum.de}}
}
\end{center}
\begin{center}
{\em Physik-Department T30, Technische Universit\"{a}t M\"{u}nchen}\\ 
{\em James-Franck-Stra{\ss}e,85748 Garching, Germany}\\ 
\end{center}

\begin{abstract}
Tools for calculating the Renormalization Group Equations
for renormalizable and non-renormalizable operators in various theories
are reviewed, which are essential for comparing experimental results 
with predictions from models beyond the Standard Model.
Numerical examples for the running of the lepton mixing angles 
in models with non-degenerate see-saw scales are shown, in which 
the best-fit values of the experimentally favored LMA solution are produced 
from maximal or from
vanishing solar neutrino mixing at the GUT scale. 
\end{abstract}

\section{Introduction}
Models for neutrino masses typically operate at high energy scales, like the GUT
scale. However from the experiments we obtain information about the low energy
values of the parameters.   
To compare them, it is essential to evolve the parameters of the models from
high to low energies. This is accomplished by the Renormalization Group
Equations (RGE's) for the operators of the theory. 
When the Standard Model (SM) or the MSSM 
is viewed as an effective field theory,
Majorana masses for the neutrinos can be introduced via an
effective operator of
mass dimension 5, which couples 2 lepton and 2 Higgs doublets.
The most promising scenarios for giving masses to neutrinos
use the see-saw mechanism, which provides a convincing
explanation for their smallness.
It can be realized by a renormalizable theory with the particle content 
of the SM or the MSSM
extended by 3 heavy neutrinos that are singlets under the SM gauge
groups. The singlets typically have large explicit 
(Majorana) masses with a 
spectrum, which is non-degenerate in general. 
Due to this non-degeneracy one has to use several effective
theories with the singlets partly integrated out, when studying the
evolution of the effective mass matrix of the light neutrinos.
We review the tools necessary to perform the Renormalization Group
analysis of the neutrino mass parameters in various models.
Numerical solutions to the RGE's show that there can be large effects for the
running of the lepton mixing angles, especially for the solar angle 
$\theta_{12}$.
The currently favored LMA solution of the solar neutrino problem 
can e.g.\ be obtained in a natural way from bimaximal mixing 
\cite{Antusch:2002hy} as well as from $\theta_{12}=\theta_{13}=0^\circ$,
$\theta_{23}=45^\circ$  \cite{Antusch:2002??} 
at the GUT scale by renormalization group effects.

\section{The Neutrino Mass Operator in the SM and in the MSSM}

\begin{figure}[h]
  $\CenterEps[1]{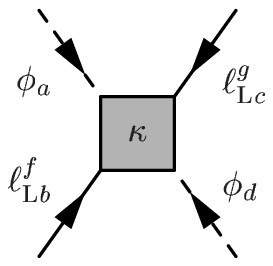}
  \xrightarrow[\mbox{{\small breaking}}]{\mbox{{\small EW symmetry}}}
  \CenterEps[1]{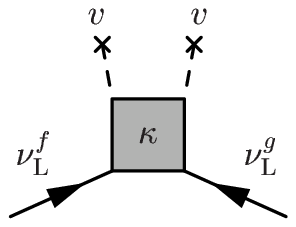}
 $
 \caption{Vertex from the dimension 5 operator, which 
 yields a Majorana mass matrix for the light
neutrinos.} \label{fig:KappaVertex}
\end{figure}

Let $\ell_{\mathrm{L}}^f$, $f\in\{1,2,3\}$, be the 
SU(2)$_\mathrm{L}$-doublets of SM leptons and
$\phi$ the Higgs doublet.
The dimension 5 operator, which gives Majorana masses to
the SM neutrinos  after electroweak (EW) symmetry breaking 
(figure \ref{fig:KappaVertex}), is given by
\begin{equation}\label{eq:Kappa:Babu:1993:1}
 \mathscr{L}_{\kappa} 
 =\frac{1}{4} 
 \kappa_{gf} \, \overline{\ell_\mathrm{L}^\mathrm{C}}^g_c\varepsilon^{cd} \phi_d\, 
 \, \ell_{\mathrm{L}b}^{f}\varepsilon^{ba}\phi_a  
  +\text{h.c.} \;.
\end{equation}
$\kappa$ is symmetric under interchange of the generation indices
 $f$ and $g$, $\varepsilon$ is the totally antisymmetric tensor in 
2 dimensions, and 
$\ell_\mathrm{L}^\mathrm{C} := (\ell_\mathrm{L})^\mathrm{C}$ is 
the charge conjugate of the lepton doublet. $a,b,c,d \in \{1,2\}$ 
are SU(2) indices. The corresponding expression in the MSSM is the $F$-term of
the part of the superpotential
\begin{eqnarray}\label{eq:Kappa-MSSM}
 \mathscr{W}_{\kappa}^{\mathrm{MSSM}} 
 =-\frac{1}{4} 
 { {\kappa}^{}_{gf}} \, {\SuperField{l}^{g}_c}\varepsilon^{cd}
 {\SuperField{h}^{(2)}_d}\, 
 \, \SuperField{l}_{b}^{f}\varepsilon^{ba} \SuperField{h}^{(2)}_a 
 +\text{h.c.} \; ,
\end{eqnarray}
where the chiral superfield  $\SuperField{l}$
contains the  lepton SU(2)$_\mathrm{L}$-doublets and $\SuperField{h}^{(2)}$
contains the Higgs doublet with weak hypercharge $+ \tfrac{1}{2}$.

\section{Calculating $\boldsymbol{\beta}$-Functions from Counterterms in MS-like Schemes}
We are interested in the \(\beta\)-function 
$ \beta_Q := \mu \frac{\D Q}{\D\mu}$
for a quantity \(Q\) in an MS-like renormalization scheme.
In general, the bare and the renormalized quantity are related by
\begin{eqnarray}
 Q_\mathrm{B} 
 & = &
 \left(\prod_{i\in I}Z_{\phi_i}^{n_i}\right)\,
 [Q+\delta Q]\mu^{D_Q\epsilon}
 \,\left(\prod_{j\in J}Z_{\phi_j}^{n_j}\right) \; ,
 \label{eq:AddRenQB}
\end{eqnarray}
where \(I=\{1,\dots, M\}\), \(J=\{M+1,\dots, N\}\), $D_Q$ is related to the
mass dimension of $Q$, $\mu$ is the renormalization scale and $\epsilon:=4-d$
stems from dimensional regularization. 
\(\delta Q\), which corresponds to the counterterm for $Q$, 
and the wavefunction renormalization constants $Z$ depend on \(Q\) and some
additional variables \(\{V_A\}\). From equation (\ref{eq:AddRenQB}), 
we obtain \cite{Antusch:2001ck}
\begin{eqnarray}\label{formula}
 \beta_Q & = & 
 \left[
      \sum_A  D_{V_A}\Braket{\frac{\D\delta Q_{,1}}{\D V_A}|V_A}
        -D_Q\,\delta Q_{,1}
         \right]
+Q\cdot\sum_{j\in J} n_j
        \left[\sum_A D_{V_A}\,\Braket{\frac{\D Z_{\phi_j,1}}{\D V_A}|V_A}\right]
	\nonumber \\
 &&+\sum_{i\in I}n_i
        \left[\sum_A D_{V_A}\,\Braket{\frac{\D Z_{\phi_i,1}}{\D V_A}|V_A}\right]
 \cdot Q \; ,
 \label{eq:BetaFunctionInAdditionalRenormalization}
\end{eqnarray}
with 
$\Braket{\frac{\D F}{\D x}|y}$ defined as
$\frac{\D F}{\D x}y$  for scalars,
$\sum_{n}\frac{\D F}{\D x_{n}} y_{n}$ for vectors, 
$\sum_{m,n}\frac{\D F}{\D x_{mn}} y_{mn}$ for matrices and analog for arbitrary
tensors $x,y$. 
The formula (\ref{formula}) can be used for any tensorial quantity $Q$.
Due to the general form of the counterterm in equation 
(\ref{eq:AddRenQB}), it can also be used for
non-multiplicative renormalization.

\section{The $\boldsymbol{\beta}$-Function for the Neutrino Mass Operator in the SM and in 2HDM's}
\begin{figure}[h]
  \begin{center}
  \subfigure[\label{subfig:f1}]{%
  \(\CenterEps[0.8]{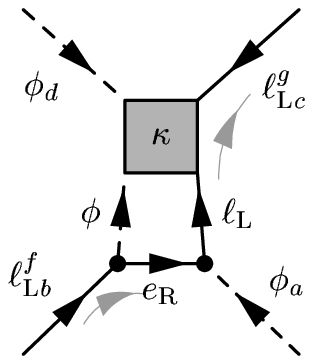}\)}
  \hfil
  \subfigure[\label{subfig:f2}]{%
  \(\CenterEps[0.8]{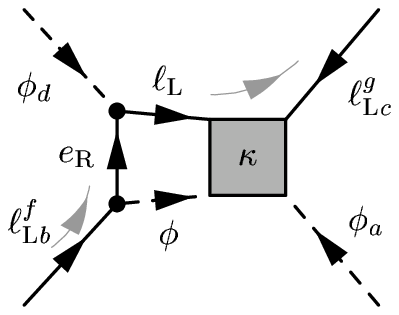}\)}
  \hfil
  \subfigure[\label{subfig:f3}]{%
  \(\CenterEps[0.8]{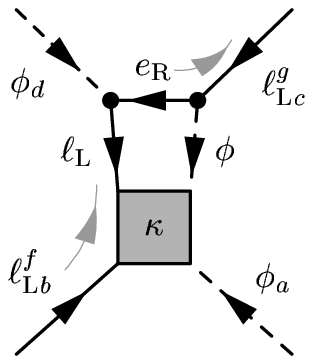}\)}
  \hfil
  \subfigure[\label{subfig:f4}]{%
  \(\CenterEps[0.8]{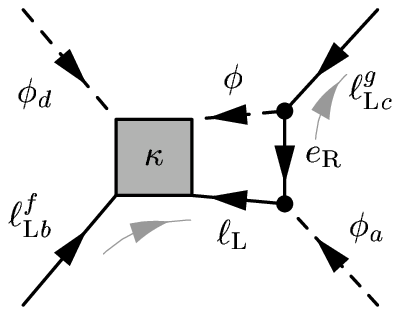}\)}
  \end{center}
 \caption{\label{fig:Non-DiagVertexDiagrams} Diagrams for the 1-loop vertex renormalization of the
 neutrino mass operator in the SM, which yield contributions to the
 $\beta$-function with a non-trivial flavour structure. 
 The gray arrow indicates the fermion flow 
as defined in \cite{Denner:1992vz}.} 
\end{figure}

Calculating the counterterm for the neutrino mass operator and the wavefunction
renormalization constants in the SM, we obtain 
for the $\beta$-function of the 
neutrino mass operator
from equation (\ref{formula}) \cite{Antusch:2001ck}
\begin{eqnarray} \label{eq:finalrge}
        16\pi^2 \beta_\kappa & = &
         -\frac{3}{2} \left[\kappa \left( Y_e^\dagger Y_e \right)
         +            \left( Y_e^\dagger Y_e \right)^T \kappa \right]
         +
        \nonumber\\
         & &
         +\lambda \kappa - 3 g_2^2 \kappa
         +2 \, \Tr \left( 3 Y_u^\dagger Y_u + 3 Y_d^\dagger Y_d 
         +Y_e^\dagger Y_e \right) \kappa \;.
\end{eqnarray}
$g_2$ is the SU(2) gauge coupling constant, $Y_u$ and 
$Y_d$ are the Yukawa matrices for the up and the down quarks, $Y_e$ is the
Yukawa matrix for the charged leptons
and $\lambda$ is the Higgs self-coupling.
Compared to earlier results \cite{Chankowski:1993tx,Babu:1993qv}, in \cite{Antusch:2001ck} we find a coefficient
$-\tfrac{3}{2}$ instead of $-\tfrac{1}{2}$ in front of the non-diagonal
term $\kappa (Y_e^\dagger Y_e) + (Y_e^\dagger Y_e)^T \kappa$ 
(see figure \ref{fig:Non-DiagVertexDiagrams}), 
which is essential for the running of the lepton mixing angles. 
Similar corrections have also been made in the RGE's for the neutrino mass
operators in Two Higgs Doublet Models (2HDM's) \cite{Antusch:2002vn}.

\section{Supergraph Construction Kit for 2-Loop $\boldsymbol{\beta}$-Functions in the MSSM}
The calculation of $\beta$-functions is simplified considerably 
in supersymmetric (SUSY) theories,
since due to the non-renormalization theorem 
\cite{Wess:1974kz,Iliopoulos:1974zv} 
only wavefunction renormalization has to be considered
for operators of the superpotential.  
However, in a component field description, 
no use can be made of the 
theorem with respect to 
gauge loop corrections since it is no longer manifest when a 
supergauge, as for example Wess-Zumino-gauge, has been fixed.
The supergraph technique 
\cite{Delbourgo:1975jg,Salam:1975pp,Fujikawa:1975ay,Grisaru:1979wc},
on the other hand, allows to use the 
non-renormalization theorem since SUSY is kept manifest. 
It can thus be used to calculate \(\beta\)-functions in 
supersymmetric theories for operators of the superpotential from the wavefunction
renormalization constants. These operators may be non-renormalizable
since for the latter the non-renormalization theorem holds as well
\cite{Weinberg:1998uv} and they do not affect the wavefunction renormalization
constants in leading order in an effective field theory expansion.
For the wavefunction renormalization constants, general formulae exist in
the literature. Thus one can formulate a construction kit for calculating 2-loop
beta functions in $N\!=\!1$ supersymmetric theories, which can be applied to
renormalizable and 
non-renormalizable operators of the superpotential \cite{Antusch:2002ek}.
We applied it to calculate the 2-loop beta functions for the
lowest-dimensional effective neutrino mass operator in the 
MSSM and for the Yukawa couplings (including $Y_\nu$) in the MSSM extended 
by singlet superfields and the Majorana mass matrix $M$ for the latter.

\section{The 2-Loop $\boldsymbol{\beta}$-Function for the Neutrino Mass Operator in the MSSM}
\begin{figure}
 \begin{center}
  \subfigure[\label{subfig:TwoLoop1}]{%
  \(\CenterEps[0.7]{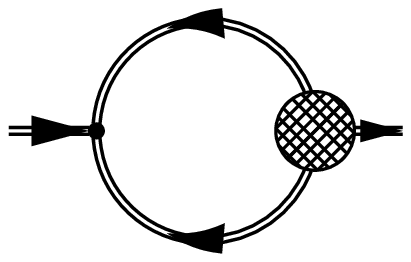}\)}
  \hfil
  \subfigure[\label{subfig:TwoLoop2}]{%
  \(\CenterEps[0.7]{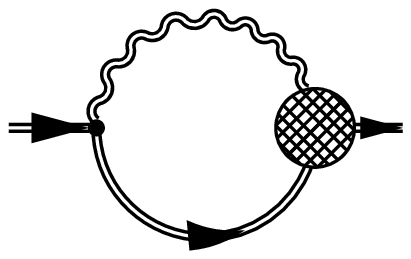}\)}
  \hfil
  \subfigure[\label{subfig:TwoLoop3}]{%
  \(\CenterEps[0.7]{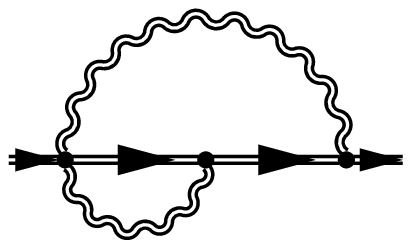}\)}
\\
  \subfigure[\label{subfig:TwoLoop4}]{%
  \(\CenterEps[0.7]{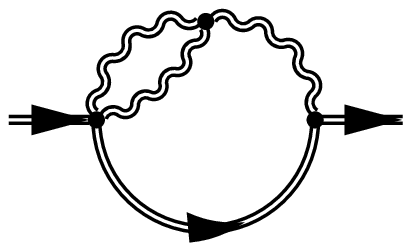}\)}
  \hfil
  \subfigure[\label{subfig:TwoLoop5}]{%
  \(\CenterEps[0.7]{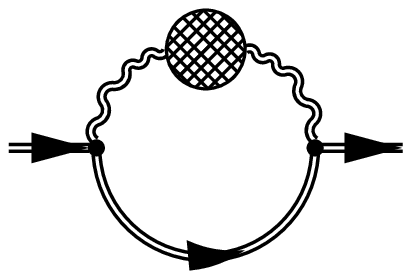}\)}
  \hfil
  \subfigure[\label{subfig:TwoLoop6}]{%
  \(\CenterEps[0.7]{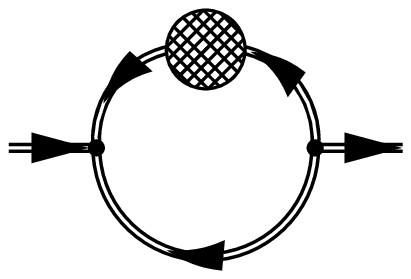}\)}
\\
  \subfigure[\label{subfig:TwoLoop7}]{%
  \(\CenterEps[0.7]{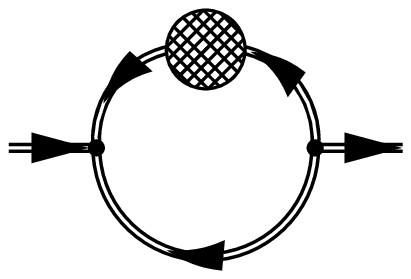}\)}
  \hfil
  \subfigure[\label{subfig:TwoLoop8}]{%
  \(\CenterEps[0.7]{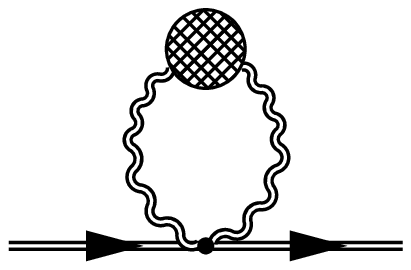}\)}
  \hfil
  \subfigure[\label{subfig:TwoLoop9}]{%
  \(\CenterEps[0.7]{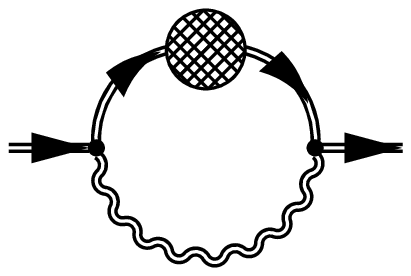}\)}
 \end{center}
 \vspace*{-0.4cm}
 \caption{2-loop supergraphs, which contribute to the 
 \(\overline{\SuperField{\Phi}}\SuperField{\Phi}\) propagator. Chiral superfields are represented as straight double
 lines while vector-superfields are indicated by wiggly double lines.
 A blob denotes
 the relevant one-particle irreducible graph including any 1-loop counterterm
 that may be required \cite{West:1984dg}.\label{fig:TwoLoopSelfEnergyDiagrams}}
\end{figure}

The calculation of the 1-loop part of the RGE for the 
neutrino mass operator in the MSSM
yields   
  \begin{equation}\label{eq:RGEMSSM}
(4\pi)^2 {\beta^{(1)}_\kappa}^{\mathrm{MSSM}}  =  
 (Y_e^\dagger Y_e)^T {\kappa}
 +{\kappa}(Y_e^\dagger Y_e)
 +6\,\Tr( Y_u^\dagger Y_u)\,{\kappa} 
 -2 g_1^2 {\kappa}- 6 g_2^2{\kappa}
 \; ,
\end{equation}
confirming the existing MSSM result \cite{Chankowski:1993tx,Babu:1993qv}.
Using the construction kit, 
from the supergraph diagrams
shown in figure \ref{fig:TwoLoopSelfEnergyDiagrams},
for the 2-loop part we obtain \cite{Antusch:2002ek}
\begin{eqnarray}
 \lefteqn{(4\pi)^4\, {\beta^{(2)}_\kappa}^{\mathrm{MSSM}} =
 \left[  -6\, \mbox{Tr}\,(Y_u^\dagger   Y_d  Y_d^\dagger  Y_u) 
 - 18\, \mbox{Tr}\,(Y_u^\dagger  Y_u  Y_u^\dagger  Y_u) 
 + \frac{8}{5}\,g_1^2\, \mbox{Tr}\,(Y_u^\dagger  Y_u)
 \right. }\nonumber \\
 &&
 + \left. 32\,g_3^2\, \mbox{Tr}\,(Y_u^\dagger  Y_u)
 + \frac{207}{25}\,\,g_1^4
 + \frac{18}{5}\,g_1^2\,g_2^2 
 + 15\,g_2^4 \right]
 \,\kappa 
 \nonumber \\
 && -
  \left[  2\, (Y_e^\dagger  Y_e  Y_e^\dagger   Y_e )^T
 -\left(
  \frac{6}{5}\,g_1^2
  -  \mbox{Tr}\,(Y_e  Y_e^\dagger) 
  - 3\, \mbox{Tr}\,(Y_d  Y_d^\dagger)
 \right)
 \,(Y_e^\dagger  Y_e)^T \right]
 \, \kappa 
 \nonumber \\
 && -\,
 \kappa \, \left[  
  2\, Y_e^\dagger  Y_e  Y_e^\dagger  Y_e 
 -\left(
  \frac{6}{5}\,g_1^2
  -  \mbox{Tr}\,(Y_e  Y_e^\dagger) 
  - 3\, \mbox{Tr}\,(Y_d  Y_d^\dagger)
 \right)
 \,Y_e^\dagger  Y_e \right] \; .
\end{eqnarray}

\vspace*{-0.1cm}

\section{Generating the LMA solution by RG Running of the Lepton Mixing Angles}
To study the RG running of the lepton mixing angles and neutrino
masses, all parameters of the
theory have to be evolved from the GUT scale to the EW
scale. Starting at
the GUT scale, the strategy is to successively solve the systems
of coupled differential equations of the form
\begin{eqnarray}
\mu \frac{\D}{\D \mu}   \accentset{(n)}{X}_i
  = \accentset{(n)}{\beta}_{{X}_i} \RaiseBrace{\Bigl(}\RaiseBrace{\Bigl\{}
  \accentset{(n)}{X}_j\RaiseBrace{\Bigl\}}\RaiseBrace{\Bigl)}
\end{eqnarray}
for all the parameters $\accentset{(n)}{X}_i \in \RaiseBrace{\bigl\{}\accentset{(n)}{\kappa},\accentset{(n)}{Y_\nu},\accentset{(n)}{M},
\dots\RaiseBrace{\bigr\}}$
of the theory.
The parameters defined in 
the energy ranges corresponding to the various effective theories are marked by
$(n)$. 
The derivation of the RGE's for the theories in the 
ranges between the see-saw scales, where the
heavy singles are partly integrated out,
and the method for dealing with these effective theories 
are given in \cite{Antusch:2002rr}.

\begin{figure}[h]
        \begin{center}
\subfigure[\label{subfig:Plot1}]{%
  \(\CenterEps[0.75]{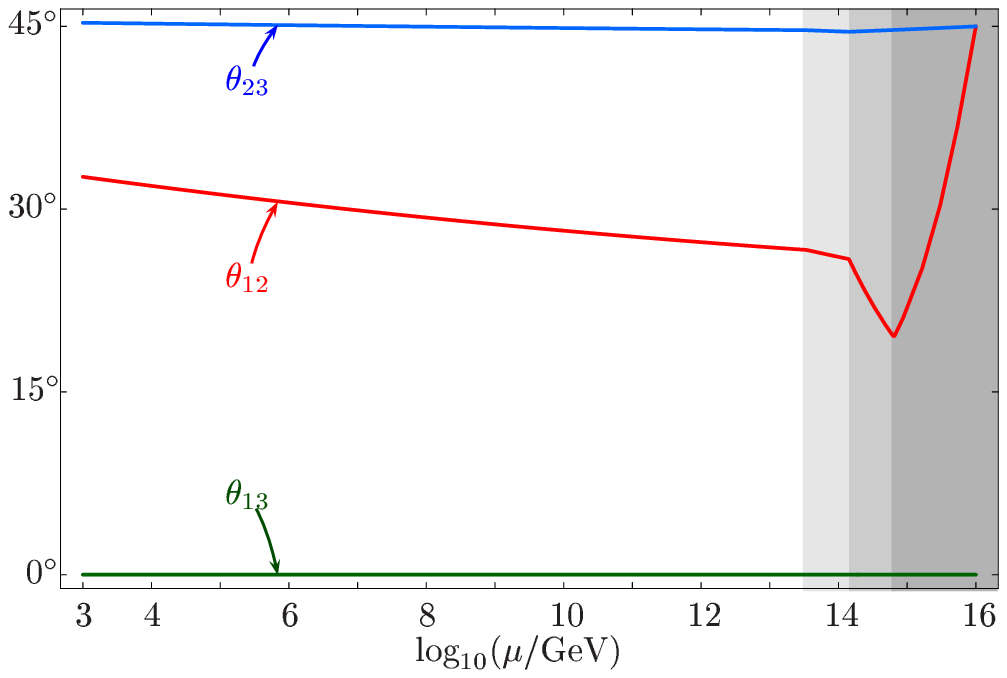}\)}
  \quad
\subfigure[\label{subfig:Plot2}]{%
  \(\CenterEps[0.75]{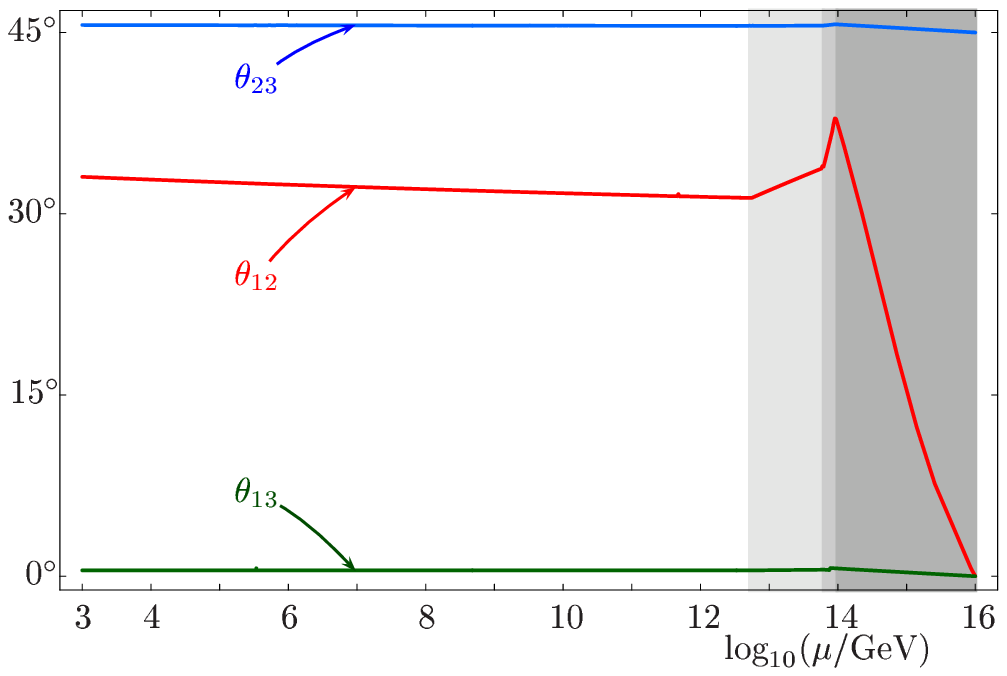}\)}              
        \end{center}
	\vspace*{-0.5cm}
\caption{\label{fig:3} 
Examples for the RG evolution of the lepton mixing angles
from the GUT scale to the SUSY-breaking scale (taken to be $\approx 1$~TeV) 
in the MSSM extended by 3 heavy singlets (right-handed neutrinos).
We assumed zero CP phases and positive mass eigenvalues for the neutrinos.
The best-fit values for the mixing angles of the LMA solution can
be obtained from bimaximal lepton mixing \cite{Antusch:2002hy}
(figure \ref{subfig:Plot1}) as well as from  
vanishing solar mixing angle at the GUT scale 
\cite{Antusch:2002??} (figure \ref{subfig:Plot2}).  
The kinks in the plots 
correspond to the mass thresholds at the see-saw scales, 
where the heavy singlets are successively integrated out. 
The grey-shaded regions mark the various effective theories between the see-saw
scales.
}
\end{figure}

The LMA solution of the
solar neutrino problem with a large but non-maximal value of the solar
mixing angle $\theta_{12}$ 
is strongly favored by the experiments.
The best-fit values are $\approx 33^\circ$ for $\theta_{12}$ \cite{Barger:2002iv,Bandyopadhyay:2002xj,Bahcall:2002hv,deHolanda:2002pp},
 $45^\circ$ for 
$\theta_{23}$ \cite{Toshito:2001dk}, while for $\theta_{13}$ at $2 \sigma$ there is an upper
bound of $\approx 9^\circ$ \cite{Apollonio:1999ae}. 
For model builders, 
especially the desired solar angle is difficult to achieve. 
This raises the question, whether the LMA
solution might be reached by RG evolution if one starts  
with bimaximal lepton mixing 
or with vanishing solar mixing (and $\theta_{13}=0^\circ,
\theta_{23}=45^\circ$) at the GUT scale. 
Figure \ref{fig:3} shows
examples for the RG evolution of the lepton mixing angles, where this has been
accomplished \cite{Antusch:2002hy,Antusch:2002??}.

\vspace*{-0.1cm}

\section{Conclusions}
The Renormalization Group analysis of the neutrino mass parameters and lepton
mixing angles provides a crucial tool towards understanding the physics at high
energy scales\footnote{A list of references to the large 
number of studies, which have investigated this
subject, can e.g.\ be found in  \cite{Antusch:2002hy}.}.
The necessary RGE's 
in see-saw scenarios for neutrino masses have been derived for the various effective
theories between the GUT scale and the low scale. For the energy ranges between
the see-saw scales, the RGE's have been derived in \cite{Antusch:2002rr}. 
The RGE's 
in the MSSM are known up to the  2-loop level \cite{Antusch:2002ek}.
This accuracy may be needed for the neutrino sector since 
due to the absence of hadronic uncertainties, high precision measurements of the
neutrino parameters may be achieved in future experiments.
The RGE for the neutrino mass operator 
at the 1-loop level has been derived in
\cite{Chankowski:1993tx,Babu:1993qv,Antusch:2001ck} for the SM, 
in \cite{Babu:1993qv,Antusch:2002vn} for 
Two Higgs Doublet Models and in
\cite{Chankowski:1993tx,Babu:1993qv,Antusch:2002vn} for the MSSM.
Numerical calculations
show that large RG evolution
of the lepton mixing angles can particularly take place 
in the energy ranges between and above the see-saw scales. 
The best-fit values for the mixing angles of the LMA solution can
be obtained from bimaximal lepton mixing \cite{Antusch:2002hy}
as well as from single maximal mixing with 
a vanishing solar mixing angle \cite{Antusch:2002??} at the GUT scale.

\section*{Acknowledgements}
The author would like to thank 
Manuel Drees, J\"orn Kersten, Manfred Lindner and Michael Ratz
for the fruitful collaboration. 
This work was supported in part by the 
``Sonderforschungsbereich~375 f\"ur Astro-Teilchenphysik der 
Deutschen Forschungsgemeinschaft''.


\begin{thebibliography}{00}


{\linespread{0.5}
\selectfont
{\small





\bibitem{Antusch:2002hy}
S.~Antusch, J.~Kersten, M.~Lindner and M.~Ratz,
\texttt{hep-ph/0206078}.

\bibitem{Antusch:2002??}
S.~Antusch and M.~Ratz,
\texttt{hep-ph/0208136}.

\bibitem{Antusch:2001ck}
S.~Antusch, M.~Drees, J.~Kersten, M.~Lindner, M.~Ratz,
 Phys. Lett.
\textbf{B519} (2001), 238--242.

\bibitem{Chankowski:1993tx}
P.~H.~Chankowski and Z.~Pluciennik,  
Phys.\ Lett.\ \textbf{B316} (1993), 312.

\bibitem{Babu:1993qv}
K.~S.~Babu, C.~N.~Leung and J.~Pantaleone,
 Phys.\ Lett.\ \textbf{B319} (1993), 191.


\bibitem{Antusch:2002vn}
S.~Antusch, M.~Drees, J.~Kersten, M.~Lindner, M.~Ratz,
Phys.\ Lett.\ \textbf{B525} (2002), 130.


\bibitem{Denner:1992vz}
A.~Denner, H.~Eck, O.~Hahn, J.~K\"ublbeck,
 Nucl.\ Phys.\ \textbf{B387} (1992), 467.

\bibitem{Wess:1974kz}
 J.~Wess and B.~Zumino, 
 Phys. Lett. \textbf{B49} (1974), 52.

\bibitem{Iliopoulos:1974zv}
 J.~Iliopoulos and B.~Zumino, 
 Nucl. Phys. \textbf{B76} (1974), 310.


\bibitem{Delbourgo:1975jg}
R.~Delbourgo,  
 Nuovo Cim. \textbf{A25} (1975), 646.

\bibitem{Salam:1975pp}
 A.~Salam and J.~Strathdee, 
 Nucl. Phys. \textbf{B86} (1975),  142--152.

\bibitem{Fujikawa:1975ay}
 K.~Fujikawa and W.~Lang,  
 Nucl. Phys. \textbf{B88} (1975), 61.

\bibitem{Grisaru:1979wc}
M.~T. Grisaru, W.~Siegel and M.~Rocek, 
Nucl. Phys. \textbf{B159} (1979), 429.

\bibitem{Weinberg:1998uv}
 S.~Weinberg,  
 Phys. Rev. Lett. \textbf{80} (1998), 3702--3705.

\bibitem{Antusch:2002ek}
S.~Antusch and M.~Ratz, 
JHEP \textbf{0207} (2002), 059.

\bibitem{West:1984dg}
P.~West, 
Phys.\ Lett.\ \textbf{B137} (1984), 371.

\bibitem{Antusch:2002rr}
S.~Antusch, J.~Kersten, M.~Lindner and M.~Ratz,
Phys.\ Lett.\ \textbf{B538} (2002), 87--95.

\bibitem{Barger:2002iv}
V.~Barger, D.~Marfatia, K.~Whisnant, and B.~P.~Wood,
Phys.\ Lett.\ \textbf{B537} (2002), 179--186.

\bibitem{Bandyopadhyay:2002xj}
A.~Bandyopadhyay, S.~Choubey, S.~Goswami and D.~P.~Roy,
Phys.\ Lett.\ \textbf{B540} (2002), 14.

\bibitem{Bahcall:2002hv}
J.~N.~Bahcall, M.~C.~Gonzalez-Garcia and C.~Pe\~na-Garay,
JHEP \textbf{0207} (2002), 054.

\bibitem{deHolanda:2002pp}
P.~C.~de Holanda and A.~Yu.~Smirnov,
\texttt{hep-ph/0205241}.

\bibitem{Toshito:2001dk}
T.~Toshito {\it et al.}\ [Super-Kamiokande Collaboration],
\texttt{hep-ex/0105023}.

\bibitem{Apollonio:1999ae}
M.~Apollonio {\it et al.}\ [CHOOZ Collaboration],
Phys.\ Lett.\ \textbf{B466} (1999), 415.


}
\par }

\end{thebibliography}
\end{document}